\documentclass[12pt]{iopart}
\usepackage{amsfonts}  
\usepackage{graphicx}
\usepackage{siunitx}
\usepackage{xcolor}
\usepackage{cite}

\newcommand{\im}{\mathrm{i}}

\newcommand{\rp}[1]{(\ref{#1})}

\newcommand{\av}[1]{\left\langle #1 \right\rangle}

\newcommand{\da}{^\dagger}

\newcommand{\pq}[1]{\left[ #1 \right]}

\newcommand{\EE}{{\cal E}}

\newcommand{\II}{{\cal I}}

\renewcommand{\figurename}{Fig.}

\newcommand{\secname}{Sec.}
\newcommand{\wc}{\omega_\mathrm{c}}
\newcommand{\ko}{\kappa_0}
\newcommand{\kp}{\kappa^{\prime}}
\newcommand{\kpp}{\kappa^{\prime\prime}}
\newcommand{\wm}{\omega_\mathrm{m}}
\newcommand{\wmj}{\omega_{\mathrm{m},j}}
\newcommand{\wml}{\omega_\mathrm{m,1}}
\newcommand{\wmr}{\omega_\mathrm{m,2}}

\newcommand{\gmj}{\gamma_{\mathrm{m},j}}
\newcommand{\gml}{\gamma_\mathrm{m,1}}

\newcommand{\dfq}{\dot q}
\newcommand{\dfp}{\dot p}
\newcommand{\dfa}{\dot a}
\newcommand{\dta}{\delta a}
\newcommand{\dtq}{\delta q}

\newcommand{\gfb}{g_\mathrm{fb}}
\newcommand{\tgfb}{\tilde g_\mathrm{fb}}
\newcommand{\ain}{a_\mathrm{in}}
\newcommand{\ainp}{a_\mathrm{in}^\prime}
\newcommand{\ainpp}{a_\mathrm{in}^{\prime\prime}}

\newcommand{\eff}{\mathrm{eff}}

\newcommand{\mct}{\mathcal{T}}

\newcommand{\tcfb}{\tilde\chi_\mathrm{fb}}
\newcommand{\as}{\alpha_\mathrm{s}}

\newcommand{\tcc}{\tilde\chi_\mathrm{c}}
\newcommand{\tccs}{\tilde\chi_\mathrm{c}^\ast}

\newcommand{\tcce}{\tilde\chi_\mathrm{c}^\mathrm{eff}}

\newcommand{\tcma}{\tilde\chi_\mathrm{m1}}
\newcommand{\tcmb}{\tilde\chi_\mathrm{m2}}
\newcommand{\tcmi}{\tilde\chi_{\mathrm{m}\,j}}

\newcommand{\tcmea}{\tilde\chi_\mathrm{m1}^\mathrm{eff}}
\newcommand{\tcmeb}{\tilde\chi_\mathrm{m2}^\mathrm{eff}}
\newcommand{\tcmei}{\tilde\chi_{\mathrm{m}\,j}^\mathrm{eff}}

\begin{document}

\title[Enhancement of three--mode optomechanical interaction by feedback--controlled light]
{Enhancement of three--mode optomechanical interaction by feedback--controlled light}

\author{Nenad Kralj$^1$, Massimiliano Rossi$^{1,2}$, Stefano Zippilli$^{1,3}$, Riccardo Natali$^{1,3}$, Antonio Borrielli$^4$, Gregory Pandraud$^5$, Enrico Serra$^{5,6}$, Giovanni Di Giuseppe$^{1,3,*}$ and David Vitali$^{1,3,\dagger}$}
\address{$^1$ School of Science and Technology, Physics Division, University of Camerino, 62032 Camerino (MC), Italy}
\address{$^2$ School of Higher Studies ``C. Urbani", University of Camerino, 62032 Camerino (MC), Italy}
\address{$^3$ INFN, Sezione di Perugia, 06123 Perugia (PG), Italy}
\address{$^4$ Institute of Materials for Electronics and Magnetism, Nanoscience-Trento-FBK Division, 38123 Povo (TN), Italy}
\address{$^5$ Delft University of Technology, Else Kooi Laboratory, 2628 Delft, The Netherlands}
\address{$^6$ Istituto Nazionale di Fisica Nucleare, TIFPA, 38123 Povo (TN), Italy}
\ead{$^*$gianni.digiuseppe@unicam.it; $^\dagger$david.vitali@unicam.it}
\vspace{10pt}
\begin{indented}
\item[]27 April 2017
\end{indented}

\begin{abstract}
We realise a feedback--controlled optical Fabry--P\'{e}rot cavity in which the transmitted cavity output is used to modulate the input amplitude fluctuations. The resulting phase--dependent fluctuations of the in--loop optical field, which may be either sub--shot or super--shot noise, can be engineered to favorably affect the optomechanical interaction with a nanomechanical membrane placed within the cavity. Here we show that in the super--shot--noise regime (``anti--squashed light'') the in--loop field has a strongly reduced effective cavity linewidth, corresponding to an increased optomechanical cooperativity. In this regime feedback improves the simultaneous resolved sideband cooling of two nearly degenerate membrane mechanical modes by one order of magnitude.
\end{abstract}

\section{Introduction}

Light fluctuations can be controlled by enclosing an optical field in a feedback loop. 
The resulting light has been studied both theoretically~\cite{WisemanMilburn2010,Jacobs2014,Taubman1995Intensity-feedb,Shapiro1987Theory-of-light,Wiseman1998In-Loop-Squeezi,Wiseman1999Squashed-states} and experimentally \cite{Buchler1999Suppression-of-,Sheard2005Experimental-de}, and in--loop field fluctuations have been shown to be either decreased (``squashed'') or increased (``anti--squashed'') for negative and positive feedback respectively. In particular, significant effort has been made to analyse the squashing regime, as a possible easy approach to the production of light with fluctuations below the vacuum noise level, i.e. of quantum squeezing. However, in--loop fields are not squeezed, even if they exhibit sub--shot--noise fluctuations; in fact, they are not free fields and as such they do not have to fulfill the standard commutation relations. As a consequence, the corresponding light extracted from the loop exhibits super--shot--noise fluctuations~\cite{Shapiro1987Theory-of-light,Wiseman1999Squashed-states}.  
Nevertheless, squashed light has been found useful for specific applications, such as
the suppression of radiation pressure noise~\cite{Buchler1999Suppression-of-}, removal of classical intensity noise \cite{Sheard2005Experimental-de}, line narrowing of atomic fluorescence \cite{Wiseman1999Squashed-states}, and more recently it has been shown that feedback--controlled light can be used to enhance optomechanical sideband cooling~\cite{Rossi2017}. Here we investigate the possibility to improve the performance of a multi--mode optomechanical system. Optomechanics~\cite{AKM-RevMod} refers to the study of the interaction of light and mechanical elements via radiation pressure, and it has emerged as a promising setting for applications in quantum technology, providing very sensitive displacement and force measurements and setting a benchmark in observing quantum--mechanical effects in macroscopic objects. Negative feedback has recently been applied in cavity optomechanics~\cite{Wilson2015Measurement-bas}. However, while in our scheme feedback is used to control the light fluctuations, in Ref.~\cite{Wilson2015Measurement-bas} light plays the role of the detector and actuator of the feedback system which operates directly on the mechanical element. 
Our approach instead shares similarities with recent ones in cavity optomechanics which make use of light with engineered fluctuations, specifically with squeezed light. Notably, optomechanical systems have been demonstrated to exhibit an improvement in both the detection sensitivity~\cite{McKenzie2002Experimental-De,LIGO2013,Peano2015Intracavity-Squ,Clark2016Observation-of-} and the cooling efficiency~\cite{Schafermeier2016aa} with the use of squeezed light. In particular, as recently observed experimentally~\cite{Clark2017} and proven theoretically~\cite{Asjad2016Suppression-of-}, the correlated fluctuations of a squeezed light field can be used to enhance optomechanical sideband cooling~\cite{AKM-RevMod,Teufel2011,Chan2011,Peterson2016Laser-Cooling-o} beyond the quantum backaction limit even in the unresolved--sideband regime. 

In our work we tailor the fluctuations of the driving field of an optomechanical system with a feedback loop, specifically, operating the system in the unusual positive feedback (``anti--squashing'') regime. We show that in this regime the response of the cavity exhibits a reduced linewidth. Correspondingly, the optomehcnical cooperativity, which measures the strength of the optomechanical interaction with respect to dissipation, is increased and the performance of the optomechanical system is improved. As already demonstrated~\cite{Rossi2017}, this scheme allows for a significant enhancement of sideband cooling of the mechanical resonator, even beyond what can be achieved with squeezed light, and in spite of the fact that the correlated in--loop fluctuations are not actually squeezed. In the present work we apply it to a system with two mechanical modes. Particularly, we improve the simultaneous cooling of the doublet $(11)$ of mechanical normal modes of a circular membrane, and show with numerical simulations that the two modes hybridise into ``bright'' and ``dark'' modes~\cite{Genes2008,Shkarin2014}.

The work is organised as follows. In \secname~\ref{sec:Theory} we give a theoretical description of the optomechanical system utilising feedback--controlled light. In \secname~\ref{sec:Experiment_Cavity} we present the experimental setup and demonstrate the effect of the feedback on the optical cavity alone. Instead, in \secname~\ref{sec:Mechanics} we show the improvement in sideband--cooling two nearly degenerate mechanical normal modes in the anti--squashing regime. We also introduce theoretically the mutual hybridisation of the two mechanical modes as mediated by the light and give simulations of this hybridisation in the present case.         

\section{Optomechanics with feedback--controlled light}
\label{sec:Theory}

\begin{figure}[h!]
\centering
\includegraphics[width=0.45\textwidth]{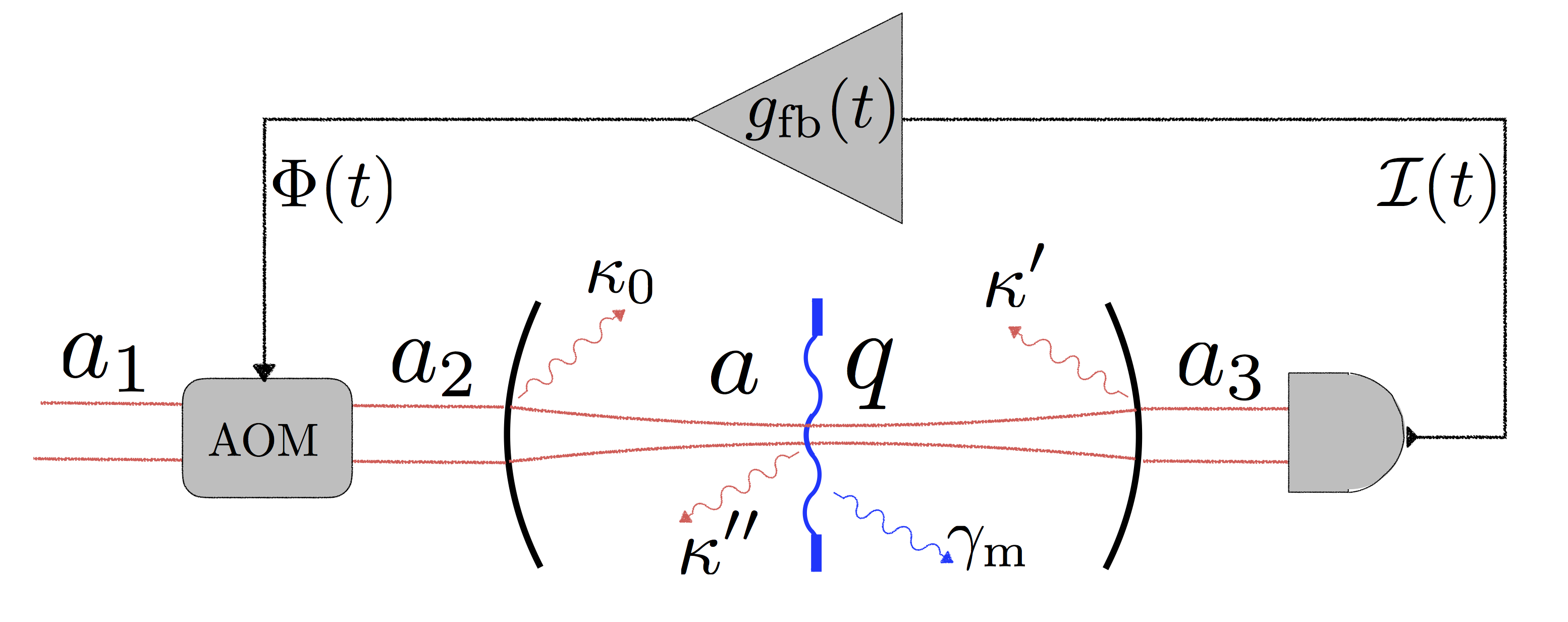}
\caption{Sketch of the dynamics. The cavity mode is described by the annihilation operator $a$ and its coupling to the environment is quantified by the decay rates $\kappa_0$ and $\kappa^\prime$ through the input and output mirror respectively, and $\kappa^{\prime\prime}$ for other losses, e.g. absorption by the membrane placed inside the cavity. This mode is also coupled to mechanical modes of the membrane, characterised by operators $q_j$ and $p_j$ and decay rates $\gamma_{\mathrm{m},j}$, via radiation pressure. The optical field transmitted by the cavity is directly detected with a photodiode. The corresponding photocurrent $\mathcal{I}(t)$ is processed electronically [encompassed by the filter function $\gfb(t)$] and the resulting signal $\Phi(t)$ is fed back to the input optical field via the amplitude--modulation port of an acousto--optic modulator (AOM).}
\label{fig:Figure_1} 
\end{figure}

We study an optomechanical system composed of a Fabry--P\'{e}rot cavity with a membrane in the middle, shown schematically in \figurename~\ref{fig:Figure_1}. We consider  
a cavity mode described by the annihilation operator $a$, with resonance frequency $\wc$, and total decay rate $\kappa = \ko + \kp + \kpp$, where $\ko$ corresponds to the losses of the input mirror, $\kp$ to those of the output mirror, and $\kpp$ to other losses; the cavity light interacts with 
two vibrational modes of the membrane described by the operators $q_j$ and $p_j$, with mechanical frequencies $\omega_{{\rm m},j}$, and mechanical decay rates $\gamma_{{\rm m},j}$, where the label $j=1,2$ is used to distinguish between the two mechanical modes. 

\noindent The equations of motion for these modes, interacting through radiation pressure, are:
\begin{eqnarray}
  \dfq_j &=\omega_{{\rm m},j}\, p_j	\label{eq:dq} \\
  \dfp_j &= -\omega_{{\rm m},j}\, q_j - \gamma_{{\rm m},j}\, p_j  + g_{0,j} a^\dag a + \xi_j \label{eq:dp}\\
  \dfa &= -(\kappa + \im \Delta_0)a + \im \sum_j\,g_{0,j}a\, q_j + \sqrt{2\ko}\,a_2\e^{-\im \theta_\Delta} 
  		+ \sqrt{2\kp}\ainp + \sqrt{2\kpp}\ainpp\label{eq:da}
\end{eqnarray}
where $g_{0,j}$ is the photon--phonon coupling between cavity light and the resonator $j$, $\Delta_0 = \wc-\omega_L$ the detuning between the cavity resonance and the driving laser frequency $\omega_L$, and the phase $\theta_\Delta = \arctan(-\Delta/\kappa)$ accounts for having chosen the phase of the cavity field as reference. Here $\Delta$ is the effective detuning, which takes into account also the optomechanical shift of the cavity resonance, the specific form of which is introduced below.
The terms $\xi_j$ are the noise operators acting on the mechanical modes, characterized by the correlation functions $\av{\xi_i(t)\,\xi_j(t')}=\gamma_{{\rm m},j}(2\,n_{{\rm th},j}+1)\delta_{ij}\delta(t-t')$, where $n_{{\rm th},j}$ is the number of thermal excitations. Instead, $a_2$, $\ainp$, and $\ainpp$ are the input noise operators for the cavity field 
corresponding to the different dissipation channels, with the latter two representing zero--average vacuum noise with $\av{\ainp(t)\,\ainp(t')\da}=\av{\ainpp(t)\,\ainpp(t')\da}=\delta(t-t')$. Conversely, the operator 
$a_2$ includes the pump field with annihilation operator $a_1=\EE+\ain$ (where $\mathcal{E} = \sqrt{\mathcal{P}/\hbar\omega_L}$ is the pump amplitude, $\mathcal{P}$ the pump power, and $\ain$ the corresponding vacuum noise), which is modulated by the feedback according to the relation
\begin{equation}\label{eq:a2}
	a_{2} = a_{1} + \Phi\, .
\end{equation}
The feedback contribution, $\Phi$, depends on the cavity field through the detection of the transmitted field
\begin{equation}
	\Phi(t)= \int dt^\prime \, \gfb(t-t^\prime)\, \mathcal{I}(t^\prime) 
\end{equation}
where $\gfb(t)$ is the causal filter function of the feedback and $\II(t)$ is the detected photocurrent 
\begin{eqnarray}
    \mathcal{I}(t)&= \left[\sqrt{\eta}\,a_3^\dag + \sqrt{1-\eta} \, c^\dag\right] \left[\sqrt{\eta}\,a_3 + \sqrt{1-\eta} \, c\right] \nonumber\\
			&=\eta\,a_3^\dag a_3 + \sqrt{\eta(1-\eta)}\,(a_3^\dag c + c^\dag a_3)	 + (1-\eta)\,c^\dag c\,,
\end{eqnarray}
with $c$ representing a vacuum field that accounts for detection inefficiency. The output field is finally given by the standard input--output relation~\cite{GardinerZoller}
\begin{equation}
	a_3 = \sqrt{2\kappa^\prime}\,a - \ainp\,.
\end{equation}
The solution of Eqs.~\rp{eq:dq}--\rp{eq:da} is usually found by the linearisation of the system for 
small fluctuations around the steady state solution, provided the system is stable, which is derived by imposing $\av{\dot{q_j}} = \av{\dot{p_j}} = \av{\dot{a}} = 0$: 
\begin{eqnarray}\label{eq:qsas}
	q_{\mathrm{s},j} = g_{0,j}\frac{\as^2}{\omega_{{\rm m},j}}
		\hspace{2cm}
	\as = \frac{\sqrt{\kappa_0}}{|\kappa +\im\Delta|}(\mathcal{E} + \bar{\Phi})\,,
\end{eqnarray}
and $p_{\mathrm{s},j} = 0$, where the choice of phase reference entails $\alpha_\mathrm{s} \in \mathbb{R}$, the effective detuning $\Delta$ including the optomechanical light--shift is
$\Delta = \Delta_0 - \sum_j\,g_{0,j}\, q_{{\rm s},j} = \Delta_0 -  \sum_j\,g_{0,j}^2\,\as^2/\omega_{{\rm m},j}$, and the averaged response of the feedback filter function is
$\bar{\Phi}=\eta\,2\kappa^\prime \as^2\int dt^\prime \gfb(t-t^\prime)$.

The linearised equations for the operators describing the fluctuations about the average values $\delta q_j=q_j-q_{{\rm s},j}$ and $\delta a=a-\as$ are (neglecting contributions at second order in the system operators)
\begin{eqnarray}
  \delta \dot q_j&=\omega_{{\rm m},j}\, p_j\\
  \dot p_j&= -\omega_{{\rm m},j}\, \dtq_j - \gamma_{{\rm m},j}\, p_j  + g_{0,j}\,\as(\dta +\dta^\dag ) + \xi_j \\
  \delta \dot{a} &= -(\kappa + \im \Delta)\,\dta + \im\,\sum_j\, g_{0,j}\, \as \,\dtq_j 
  		+ \sqrt{2\kappa_0}\,\delta \Phi_{\rm a}\,\e^{-\im\theta_\Delta} +
				\nonumber \\ & \hspace{1.5cm}
		+ \sqrt{2\kappa_0}\,(\delta \Phi_{\rm n} + \ain)\,\e^{-\im\theta_\Delta}
  		+ \sqrt{2\kp}\,\ainp  + \sqrt{2\kpp}\,\ainpp		\,,
\end{eqnarray}
where $\Phi$ in Eq.~\rp{eq:a2} was decomposed as $ \Phi = \bar{\Phi} +  \delta \Phi_{\rm a} +  \delta \Phi_{\rm n}$ with
\begin{eqnarray}
    \delta \Phi_{\rm a} &= 
    		\eta\,2\kp\,\as \int d t^\prime \gfb(t-t^\prime)\left[\dta(t^\prime) + \dta^\dag(t^\prime) \right] \\
    \delta \Phi_{\rm n} &= 
    		-\eta\,\sqrt{2\kp}\,\as \int d t^\prime \gfb(t-t^\prime)\left[\ainp(t^\prime) 
						+a_\mathrm{in}^{\prime\,\dag}(t^\prime) \right] +
							\nonumber\\&\hspace{1.5cm}
						+ \sqrt{\eta(1-\eta)}\,\sqrt{2\kp}\,\as\int d t^\prime \gfb(t-t^\prime)\left[c(t^\prime) 
						+  c^\dag(t^\prime)\right]\,.
\end{eqnarray}
Eliminating the momentum operator and introducing the equation for $\dta^\dag$, we have
\begin{eqnarray}
  \delta \dot{a} &= -(\kappa + \im \Delta)\,\dta + \im\sum_j\, g_{0,j}\, \as \,\dtq_j 
  		+ \sqrt{2\ko}\,\delta \Phi_{\rm a}\,\e^{-\im\theta_\Delta} +
							\nonumber\\&\hspace{2.75cm}
		+ \sqrt{2\ko}\,(\delta \Phi_{\rm n} + \ain)\,\e^{-\im\theta_\Delta} \label{eq:dta}
  		+ \sqrt{2\kp}\,\ainp  
		+ \sqrt{2\kpp}\,\ainpp\\
  \delta \dot{a}^\dag &= -(\kappa - \im \Delta)\,\dta^\dag - \im\sum_j g_{0,j}\, \as \,\dtq_j
  		+ \sqrt{2\ko}\,\delta \Phi_{\rm a}\,\e^{\im\theta_\Delta}	 +		 \label{eq:dtadag}  
							\nonumber\\&\hspace{2.75cm}		
		+ \sqrt{2\ko}\,(\delta \Phi_{\rm n} + a_\mathrm{in}^\dag)\,\e^{\im\theta_\Delta} 
		+ \sqrt{2\kp}\,a_\mathrm{in}^{{\prime\,^\dag}}
		+ \sqrt{2\kpp}\,a_\mathrm{in}^{{\prime\prime\,^\dag}}\\
  \delta \ddot q_j&= -\omega_{{\rm m},j}^2 \dtq_j - \gamma_{{\rm m},j}\,\delta \dot q_j  
  			+ g_{0,j}\,\as\,\omega_{{\rm m},j}(\dta + \dta^\dag ) 
			+\omega_{{\rm m},j}\, \xi_j 	\label{eq:dtq}\,.
\end{eqnarray}
In the frequency domain Eqs.~\rp{eq:dta}--\rp{eq:dtq} become
\begin{eqnarray}\label{eq:da_omega}
  -\im\omega\,\delta \tilde{a} &= -(\kappa + \im \Delta)\,\delta \tilde{a} + \im\sum_j\, g_{0,j}\, \as \,\delta \tilde{q}_j\, 
  		+ \sqrt{2\ko}\,\delta \tilde{\Phi}_{\rm a}\,\e^{-\im\theta_\Delta} + \tilde{n}\\
  -\im\omega\,\delta \tilde{a}^\dag &= -(\kappa - \im \Delta)\,\delta \tilde{a}^\dag - \im\sum_j\, g_{0,j} \as\,\delta \tilde{q}_j\,
  		+ \sqrt{2\ko}\,\delta \tilde{\Phi}_{\rm a}\,\e^{\im\theta_\Delta} +  \tilde{n}^\dag
  		\label{eq:dad_omega}
  		\\
  -\omega^2 \delta \tilde{q}_j&= -\omega_{{\rm m},j}^2 \delta \tilde{q}_j +\im\omega\, \gamma_{{\rm m},j}\,\delta \tilde{q}_j  
  			+ g_{0,j}\,\omega_{{\rm m},j}\,\as(\delta \tilde{a} + \delta \tilde{a}^\dag ) 
		+\omega_{{\rm m},j}\, \tilde{\xi}_j\,,\label{eq:dq_omega}
\end{eqnarray}
where the symbol $\delta \tilde{a}^\dag$ does not indicate the Hermitian conjugate of $\delta \tilde{a}$, rather the Fourier transform of $\delta a\da$, so that $\pq{\delta \tilde{a}(\omega)}\da=\delta \tilde{a}\da(-\omega)$ (and similar for all the other operators in frequency domain); moreover
\begin{eqnarray}
	\tilde{n} &= \sqrt{2\ko}\,(\delta  \tilde{\Phi}_{\rm n} + \tilde{a}_{\mathrm{in}})\e^{-\im\theta_\Delta} 
  		+ \sqrt{2\kp}\,\tilde{a}_{\mathrm{in}}^\prime + \sqrt{2\kpp}\,\tilde{a}_{\mathrm{in}}^{\prime\prime}
\end{eqnarray}
and
\begin{eqnarray}
    \delta  \tilde{\Phi}_{\rm a} &= \eta\,2\kp\, \tgfb(\omega)\,\as\left(\delta \tilde{a} 
				+ \delta \tilde{a}^\dag \right)\\
    \delta \tilde{\Phi}_{\rm n} &= -\eta\,\sqrt{2\kp}\,\tgfb(\omega)\,\as\left(\tilde{a}_\mathrm{in}^\prime 
						+ \tilde{a}_\mathrm{in}^{\prime\,\dag} \right) \,+
							\nonumber\\&\hspace{4.75cm}
			+ \sqrt{\eta(1-\eta)}\sqrt{2\kappa^\prime}\, \tilde{g}_{\rm fb}(\omega)\,\as\left(\tilde{c} 
						+   \tilde{c}^\dag\right) 	\,,				
\end{eqnarray}
with $\tgfb^\ast(-\omega)= \tgfb(\omega)$. 

Let us first inspect the effects of the feedback on the cavity field alone, assuming for the time being that $g_{0,j} = 0$. This can experimentally be realised by placing the membrane exactly at a node or an anti--node of the cavity field. The cavity field operator, obtained by solving Eqs.~\rp{eq:da_omega}--\rp{eq:dad_omega}, then takes the form
\begin{eqnarray}
	\delta\tilde{a} = 
	\tcce(\omega)
		\Big[ \tilde{n} -	\tcfb(\omega) \,\tcc(-\omega)^*
		\left(\tilde{n}\,\e^{\im\theta_\Delta} - \tilde{n}^\dag\,\e^{-\im\theta_\Delta}\right)\Big]\,,
\end{eqnarray}
where we have introduced the feedback--modified cavity susceptibility 
\begin{equation}
	\tcce(\omega) = \frac{\tcc(\omega)}
		{1 - \tcfb(\omega)\Big[ \tcc(\omega)\,\e^{-\im \theta_\Delta}
				+ \tccs(-\omega)\,\e^{\im \theta_\Delta}\Big]}\,,
\label{eq:ChiEff}
\end{equation}
with $\tcc(\omega) = [\kappa {-} \im (\omega - \Delta)]^{-1}$
the bare cavity susceptibility, and $\tcfb(\omega) =\eta \sqrt{2\ko}\,2\kp\,\as \,\tgfb(\omega)$. 
We further note that the condition for feedback stability is given in the frequency domain by $\mathrm{Re}\{\tcfb(\omega)[ \tcc(\omega)\,\e^{-\im \theta_\Delta}+ \tccs(-\omega)\,\e^{\im \theta_\Delta}]\}\leq1$. 

In order to study the response of the cavity with feedback, we inject a seed from the input mirror -- larger than all noise terms, but still small enough not to change the mean amplitude -- such that $\tilde n(\omega) \sim \sqrt{2\ko}\,\e^{-\im\theta_\Delta}\,\tilde\alpha_\mathrm{seed}$ and $\tilde n^\dag(\omega) \sim \sqrt{2\ko}\,\e^{\im\theta_\Delta}\,\tilde\alpha_\mathrm{seed}$. In this case the transmitted field becomes 
\begin{eqnarray}
	\tilde a_3 \sim \sqrt{2\kp}\tilde \dta
			     \sim 2\sqrt{\ko\kp}\,\tcce(\omega)\,\e^{-\im\theta_\Delta}\,\tilde\alpha_\mathrm{seed}\,,
\end{eqnarray}
and the transmission coefficient, defined as $\tilde t(\omega) \equiv \tilde a_3/\tilde \alpha_\mathrm{seed}$, is in turn given by
\begin{equation}
	\tilde t(\omega) = 2\sqrt{\ko\kp}\,\e^{-\im\theta_\Delta}\,\tcce(\omega) \,.
\label{eq:transmCoeffGen}
\end{equation}
As a particular case of interest, for $\Delta \gg \kappa$, we have for frequencies close to cavity resonance ($\omega \sim \Delta$) that
$\tcce(\omega) \sim
	[\kappa_{\mathrm{eff}}-\im(\omega - \Delta_{\mathrm{eff}})]^{-1}$,
where $\kappa_\eff = \kappa \left\{ 1 - \mathrm{Re} \left[ \mct(\Delta) \right] \right\}\,$, $\Delta_\eff = \Delta - \kappa \, \mathrm{Im}\left[ \mct(\Delta) \right]$ and $\mct(\omega) = \tcfb(\omega)\,\tcc(\omega)\,\e^{-\im \theta_\Delta}$ is the complete open--loop transfer function.
The transmission coefficient becomes, accordingly,
\begin{equation}
	\tilde t(\omega) \sim \frac{2\sqrt{\ko\kp}\,\e^{-\im\theta_\Delta}}{\kappa_\eff - \im( \omega - \Delta_\eff )} \, .
\label{eq:transmCoeff}	
\end{equation}

\section{Experimental setup and cavity response}
\label{sec:Experiment_Cavity}

\begin{figure}[h!]
\centering
\includegraphics[width=0.65\textwidth]{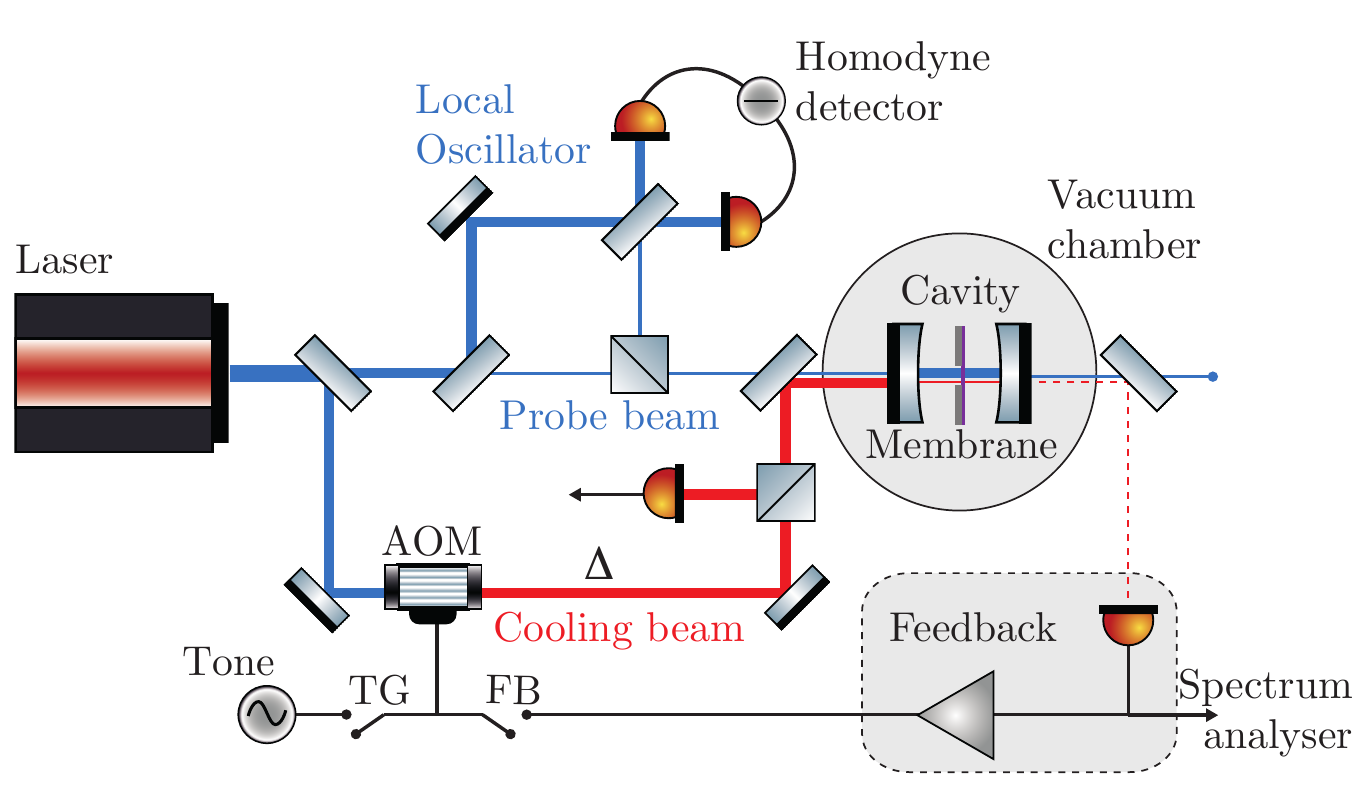}
\caption{\label{fig:Figure_2} Scheme of the experiment. The \SI{1064}{\nano\meter} laser generates two beams. The \textit{probe} beam (blue lines) is used to monitor the cavity frequency fluctuations and to lock the laser frequency to a cavity resonance. The \textit{cooling} beam (red lines) is used to drive the cavity mode and it is the field on which the feedback operates. Its frequency is shifted by means of an acousto--optic modulator (AOM). The amplitude quadrature of the transmitted light is measured via direct detection after being filtered by the cavity. The resulting photocurrent is filtered electronically and the output signal is finally used to modulate the input field amplitude, via the amplitude--modulation port of the same AOM. The TG and FB switches enable the measurement of the open-- and closed--loop transfer functions of the scheme.}
\end{figure}

The scheme of the experiment is shown in 
\figurename~\ref{fig:Figure_2}.
Two beams are derived from a \SI{1064}{\nano\meter} master laser. The \textit{probe} beam (blue lines) is used to lock the laser frequency to the cavity resonance via the Pound--Drever--Hall (PDH) technique \cite{Drever1983}. The phase fluctuations of the cavity are measured by monitoring the phase of the reflected probe field via homodyne detection \cite{Yuen1983}.
The \textit{cooling} beam (red lines) is the one on which the feedback is applied. Two cascaded acousto--optic modulators (indicated in the figure only as AOM) shift its frequency with respect to the cavity resonance, introducing a detuning $\Delta$~\cite{Karuza2013}. Owing to their orthogonal polarisations, the two transmitted beams are split with a high extinction ratio. The amplitude quadrature of the transmitted cooling field is directly detected with a single InGaAs photodiode. The generated photocurrent is then converted into a voltage signal by means of a transimpedance amplifier and filtered by a proportional--derivative controller with a corner frequency of \SI{150}{\kilo\hertz} (indicated in the figure as Feedback). The resulting electronic signal is applied to the AOM to modulate the amplitude of the input field.

\begin{figure}[h!]
\centering
\includegraphics[width=0.75\textwidth]{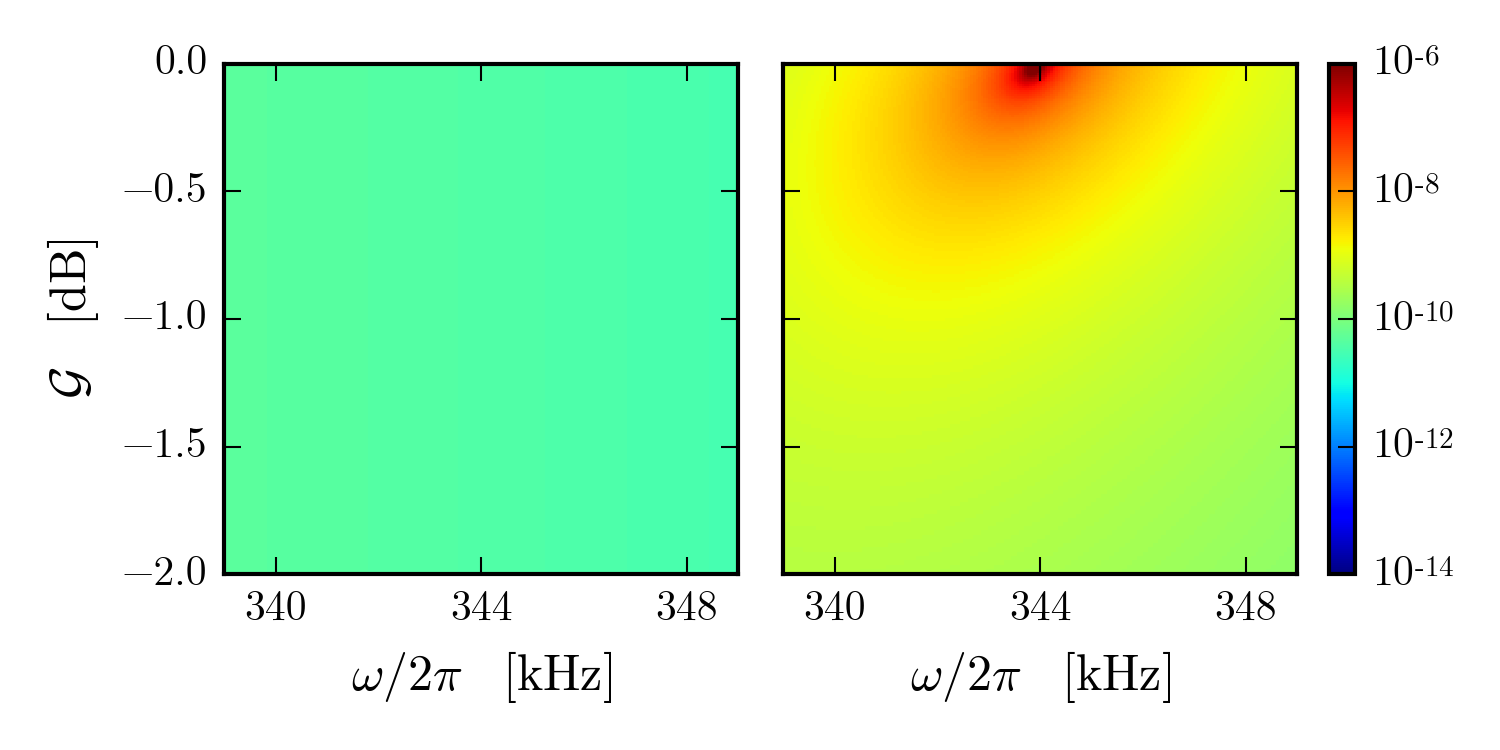}
\caption{Magnitude of the cavity response without (Left) and with (Right) the feedback--loop for a fixed detuning $\Delta = 2\pi\times\SI{330}{\kilo\hertz}$ and positive gain, i.e. in the anti--squashing regime.}
\label{fig:Figure_4} 
\end{figure}

\begin{figure}[h!]
\centering
\includegraphics[width=0.75\textwidth]{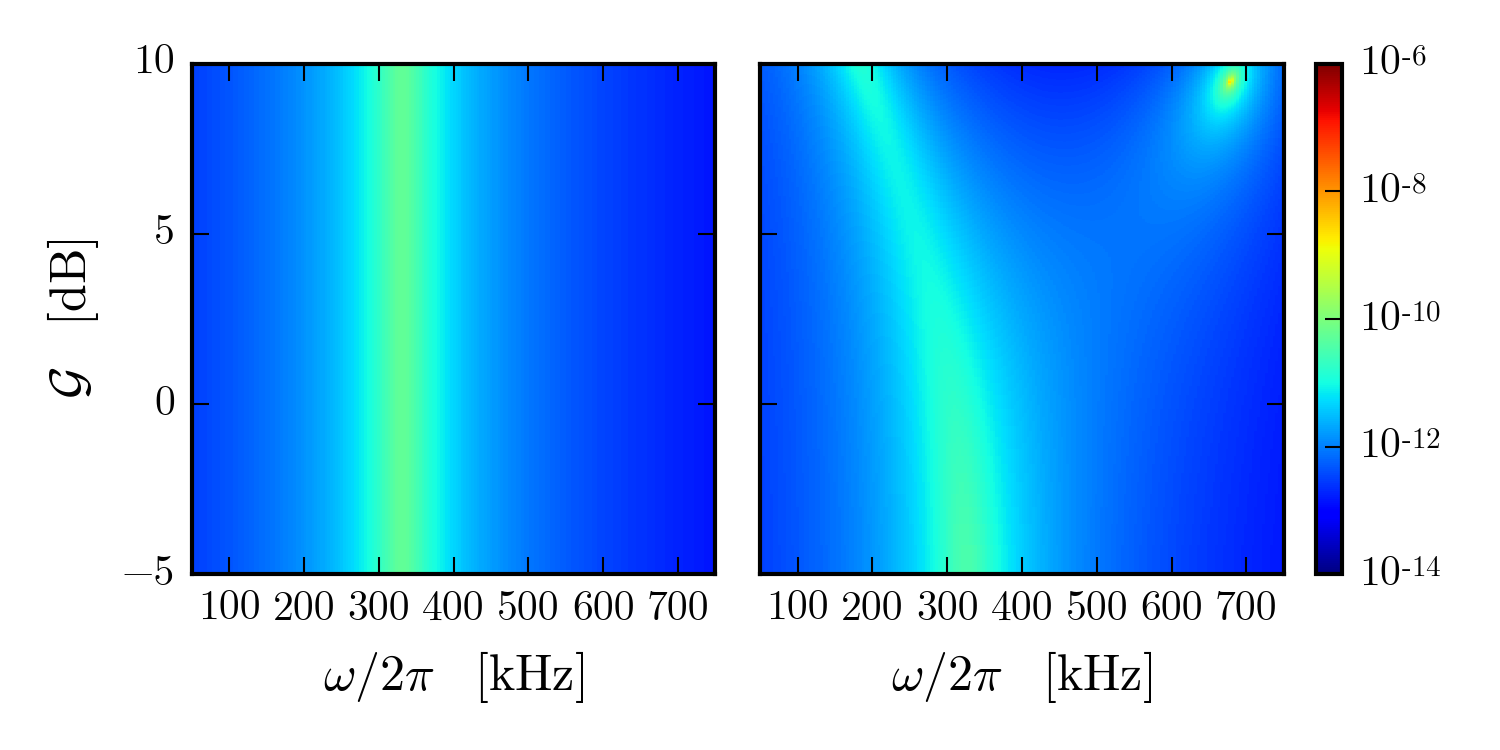}
\caption{Magnitude of the cavity response without (Left) and with (Right) the feedback--loop for a fixed detuning $\Delta = 2\pi\times\SI{330}{\kilo\hertz}$ and negative gain, i.e. the squashing regime.}
\label{fig:Figure_5} 
\end{figure}

To have a complete understanding of the in--loop modification of light fluctuations, the membrane was placed at a node or an anti--node of the cavity standing wave, reproducing the condition $g_{0,j} = 0$~\cite{Biancofiore2011}. The feedback loop was first characterised by finding its open--loop transfer function, for which purpose the TG switch of the figure was closed and the tracking generator of the network analyser provides a frequency swept tone to the AOM input, producing the seed on the cooling beam. 
\begin{figure}[h!]
\centering
\includegraphics[width=0.75\textwidth]{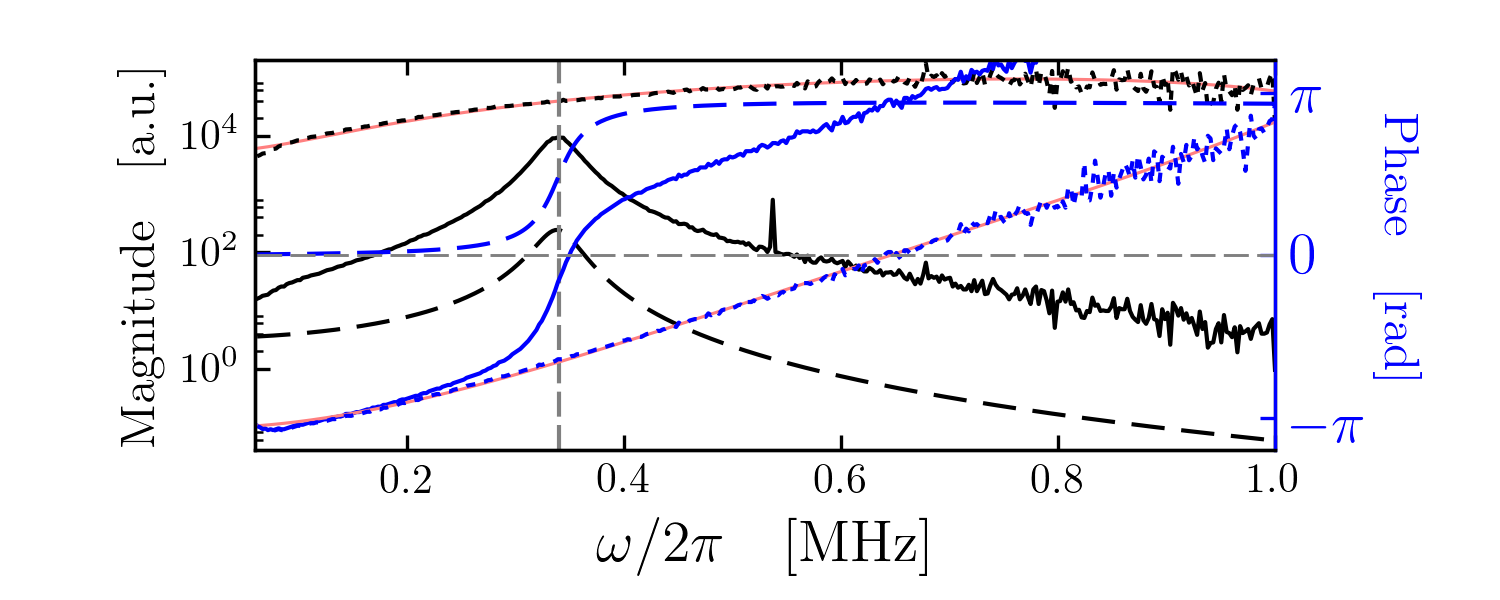}
\caption{The feedback loop of the transmitted light contains both optical and electronic filters, such that the complete measured open--loop response $\mathcal{T}(\omega)$ contains both the optical and the electronic transfer function. Black and blue lines correspond to the magnitude and phase of the three complex functions, respectively. Solid lines are the data measured; dashed lines are estimates of the cavity transfer function; dotted lines, pertaining to the electronic filter, are obtained by dividing the data by the cavity function. The vertical dashed grey line represents the detuning of the cooling beam. Light--red lines are best fits of the electronic transfer function assuming a fourth--order polynomial.}
\label{fig:Figure_3} 
\end{figure}
The electronic signal is analysed with the switch FB open, i.e. after the feedback: comparing the magnitude and phase to those of the input tone yields the complete open--loop transfer function $\mathcal{T}(\omega)$. In particular, as shown separately for the magnitude and phase in \figurename~\ref{fig:Figure_3}, for a proper characterisation of the feedback the measured open--loop response (full lines) is divided into the cavity (dashed lines) and the electronic (dotted lines) transfer function. The cavity part is determined by $\tcc(\omega)$ and is estimated from the known detuning $\Delta$ and the ``bare'' cavity linewidth measured with a ringdown technique to be $\kappa = 2\pi \times \SI{20.1}{\kilo\hertz}$. Dividing one by the other leaves the electronic filter transfer function, the slope of the phase of which can be used to estimate the feedback delay--time as $\tau_\mathrm{fb} \approx \SI{750}{\nano\second}$. However, the system turns out to be extremely sensitive to the filter phase, which is why the following simulations and fits of the cavity response make use of the best fit of the filter phase (and the corresponding magnitude) with a fourth--order polynomial, indicated in \figurename~\ref{fig:Figure_3} by light--red lines.      
Having characterised the feedback loop, we proceed to study the effective in--loop cavity susceptibility. 
The magnitude of the cavity response as a function of frequency, provided by the theoretical expression in Eq. (\ref{eq:ChiEff}) in conjunction with the fourth--order polynomial fit of the filter transfer function, is simulated in Figs.~\ref{fig:Figure_4} and \ref{fig:Figure_5} with and without the feedback loop. 

The detuning of the cooling beam is fixed at $\Delta = 2\pi\times\SI{330}{\kilo\hertz}$, while the feedback gain is positive (anti--squashing regime) for the former figure and negative (squashing regime) for the latter, and it is generally expressed in terms of a normalised factor $\mathcal{G} = \mathrm{Re}\left[ \mathcal{T}(\Delta) \right]$, such that it takes the value $\mathcal{G} = 1$ at the stability threshold, i.e. when $\kappa_\eff = 0$. Consequently, this is the upper bound of the gain in the anti--squashing regime, towards which the cavity response exhibits a sharp peak. On the other hand, in the squashing regime the cavity linewidth increases and a second peak starts to appear due to the phase change imposed by the feedback delay--time. 

\begin{figure}[h!]
\centering
\includegraphics[width=0.75\textwidth]{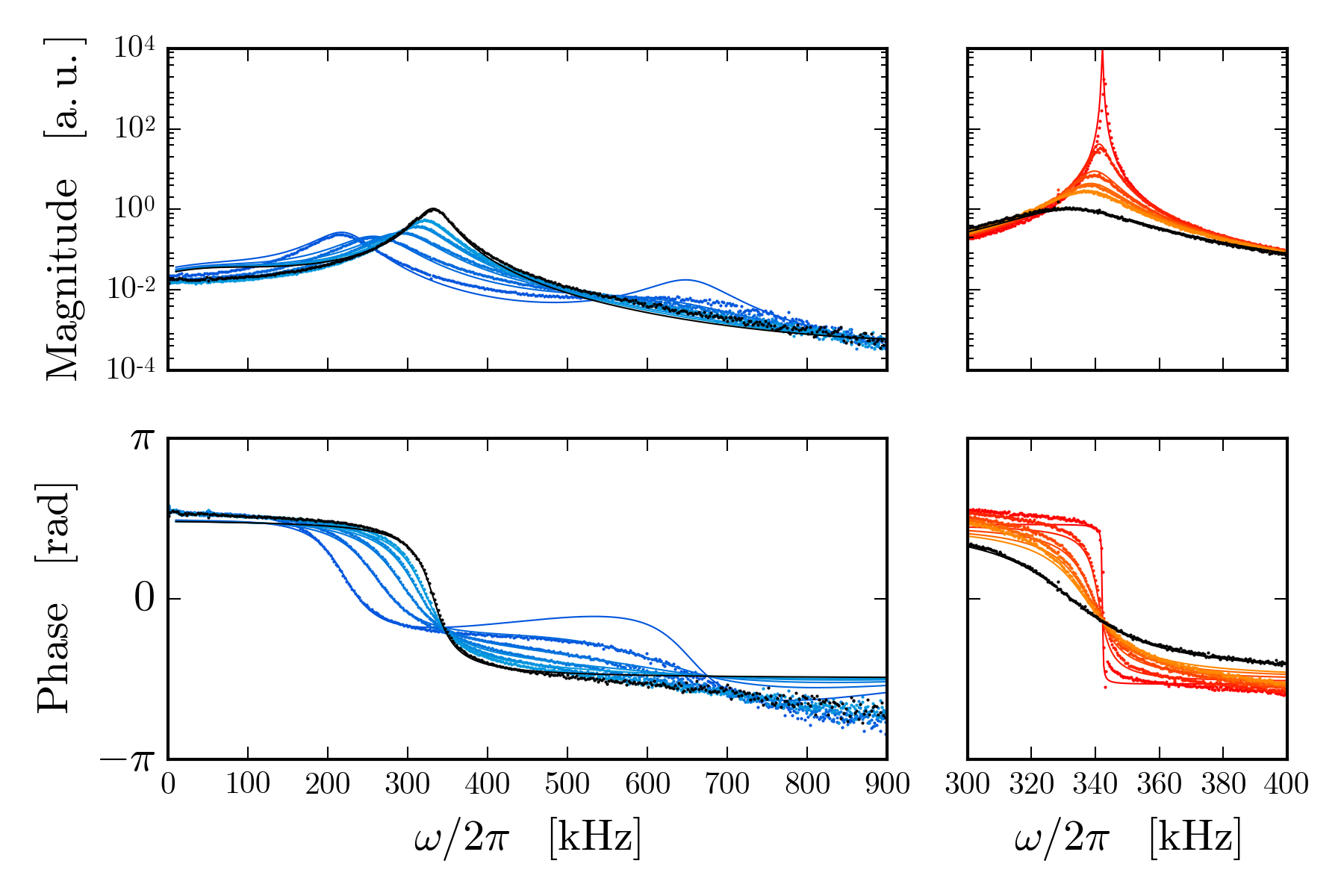}
\caption{Magnitude and phase of the transmitted seed for the squashing (Left) and anti--squashing regime (Right) for a fixed detuning $\Delta = 2\pi\times\SI{330}{\kilo\hertz}$. In the case of squashing the feedback gain is in the range $[0.5,10]$ (from light-- to dark--blue); in the opposite regime the gain is in the range $[.2,.99]$ (from orange to red). Dots represent data, while curves are theoretical expectations. The black curve represents the transmission spectrum without feedback.}
\label{fig:Figure_6} 
\end{figure}

We experimentally obtain this effective in--loop cavity susceptibility from the closed--loop transfer function, i.e. by closing both the TG and the FB switch in \figurename~\ref{fig:Figure_2}. From the transmitted cooling field, we detect the corresponding frequency component, measuring its amplitude and phase shift with respect to the injected seed. By scanning the frequency around the cavity resonance (which is at $\Delta$ with respect to the cooling beam), the complete cavity susceptibility is revealed, as reported in \figurename~\ref{fig:Figure_6} for positive (orange to red) and negative (light-- to dark--blue) feedback gain. 
As predicted by the simulation, in the former case the magnitude exhibits a narrowing of the cavity linewidth $\kappa_\eff$, up to a minimum of $\kappa_\eff^{\mathrm{min}} \approx \SI{250}{\hertz}$. On a wider scale, the latter shows an emergence of other peaks due to the loop delay--time, as confirmed by the corresponding change in phase.

\begin{figure}[h!]
\centering
\includegraphics[width=0.675\textwidth]{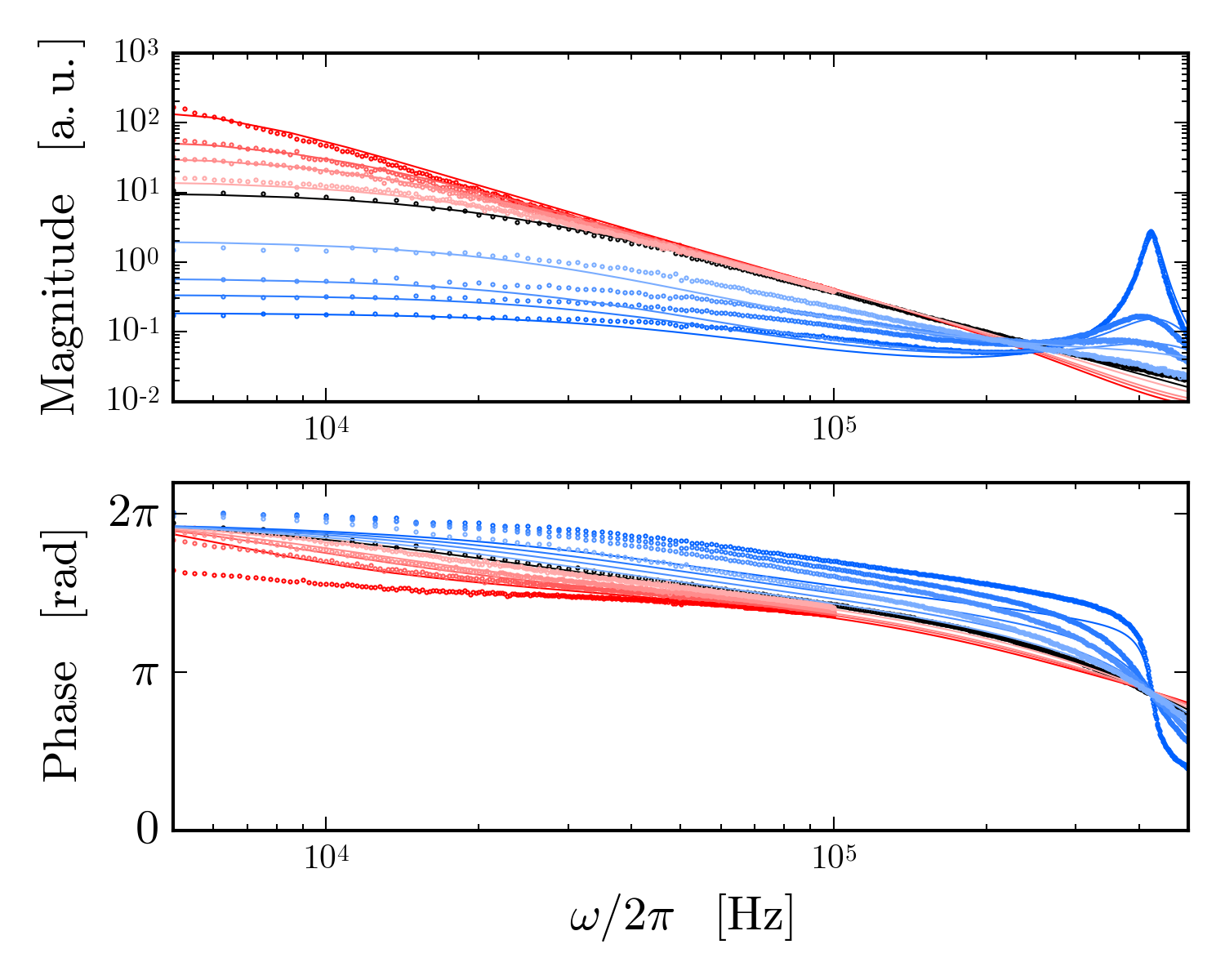}
\caption{Magnitude and phase of the transmitted seed for detuning $\Delta = 0$. In the case of squashing the feedback gain is $[1.0,2.5,4.5,7.5]$ (from light-- to dark--blue); in the opposite regime the gain is $[0.23,0.59,0.77,0.99]$ (from light-- to dark--red). Dots represent data, while curves are theoretical expectations. The black curve represents the transmission spectrum without feedback.}
\label{fig:Figure_7} 
\end{figure}

From an optomechanical standpoint, in our setup the limit given in Eq.~(\ref{eq:transmCoeff}) is useful (for frequencies around $\Delta$) because it is within the resolved--sideband regime, $\wm \gg \kappa$, in which the optimal detuning for sideband cooling is $\Delta = \wm$, and therefore $\Delta \gg \kappa$. Conversely, some optomechanical systems are operated in the unresolved--sideband regime, i.e. $\wm \ll \kappa$, in which case the minimum phonon occupancy is reached for $\Delta = \kappa/2$. The latter systems can also benefit from the feedback scheme presented herein, since the effective cavity linewidth can be reduced irrespective of the ratio $\Delta/\kappa$. We corroborate this claim by demonstrating the cavity response for the cooling beam detuning set at $\Delta = 0$.    
The corresponding magnitude and phase of the transmitted seed are shown in~\figurename~\ref{fig:Figure_7}, with the results for both feedback regimes fitted over the same frequency range.
Again it is apparent that for anti--squashing of light (light-- to dark--red) the cavity response is sharpened with respect to that with no feedback (black trace). For the squashing regime the situation is reversed and the peak to the right is once again due to a finite $\tau_{\mathrm{fb}}$.

\section{Mechanical displacement}
\label{sec:Mechanics}

We now move on to consider the interaction of light with the mechanical resonator, i.e. the membrane, arising when the membrane is moved to a position where the optomechanical coupling is significant. In the high--temperature regime, relevant to our experiment, the figure of merit which determines the sideband cooling efficiency is the cooperativity parameter $\mathcal{C}_j = 4 \, g_{0,j}^2 \, \as^2/\kappa \, \gamma_{\mathrm{m},j}$, given by the ratio between the optomechanical coherent interaction strength and the decay rates of the modes~\cite{AKM-RevMod}. This observation suggests that the reduced cavity linewidth obtained with feedback can be exploited to efficiently enhance the cooling performance.  

\begin{figure}[h!]
\centering
\includegraphics[width=0.65\textwidth]{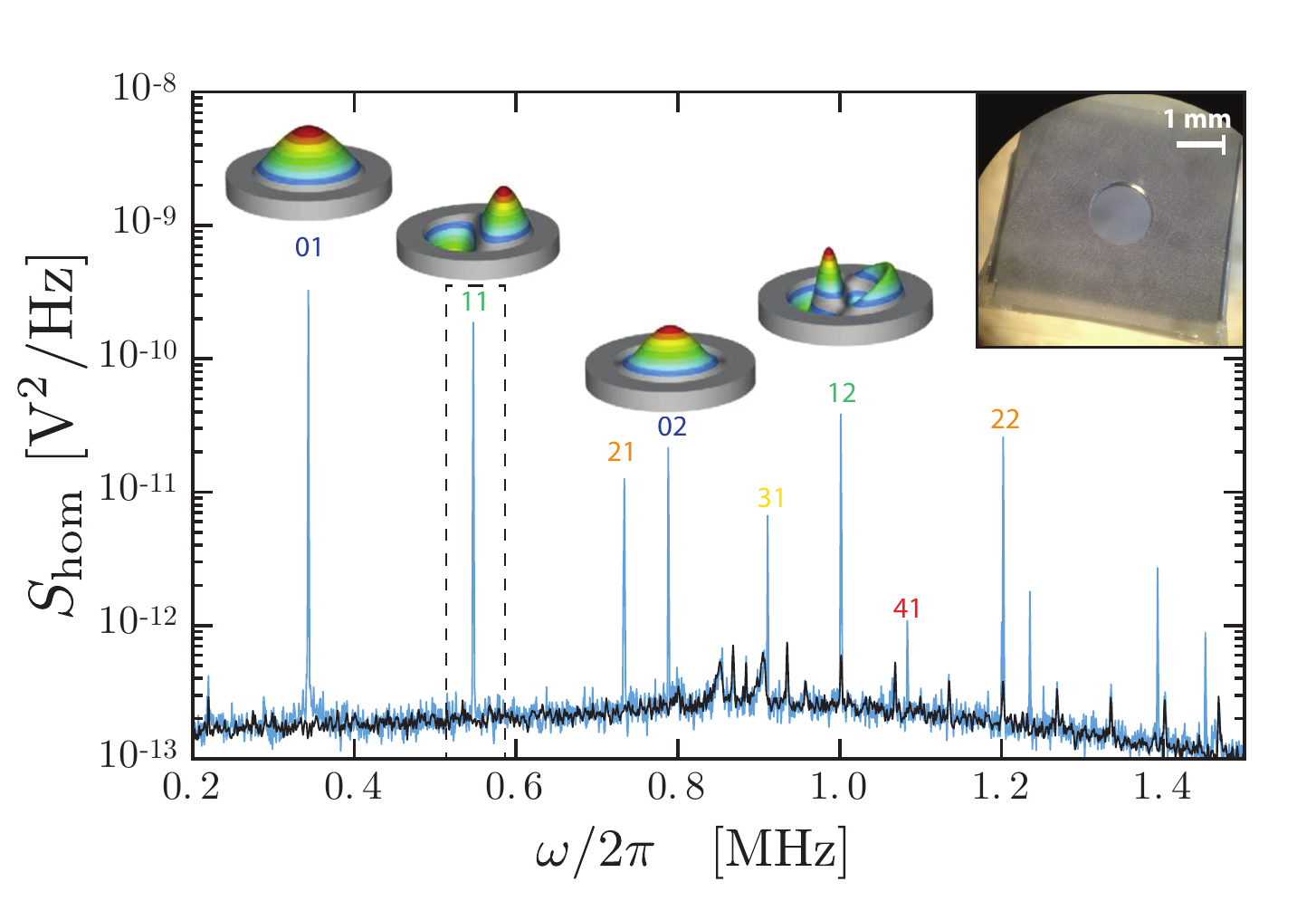}
\caption{\label{fig:Figure_8} Spectrum of the phase quadrature of the reflected resonant probe beam, measured via homodyne detection. The black trace is the detection shot noise. The signal, shown in blue, bears all cavity fluctuations, with most of the peaks due to the membrane thermal motion, which can be decomposed in a set of normal modes. The SiN membrane, presented in the upper right corner, is a circular one, \SI{1.2}{\milli\meter} in diameter and \SI{97}{\nano\meter} in thickness. The normal mode corresponding to each peak is indicated by the labels above. In particular, four transverse deformation functions are pictorially represented above the matching modes. The highlighted mode is the (11), actually a doublet because of broken cylindrical symmetry.}
\end{figure}

In this section we study a circular SiN membrane, \SI{97}{\nano\meter} thick and \SI{1.2}{\milli\meter} in diameter~\cite{Serra2016}, under the effect of a feedback--controlled pump (cooling) field, in terms of corresponding displacement spectra given by the homodyne photocurrent of the resonant probe beam.
The spectrum in \figurename~\ref{fig:Figure_8}, obtained with the cooling beam off, is used to characterise the membrane with regard to its normal modes. 
The noise floor (black trace) is the detection shot noise, which is \SI{12}{\decibel} above the electronic noise. The resonant condition of the probe, in addition to its small power of \SI{25}{\micro\watt}, guarantees that the measured mechanical displacement is not affected by optical forces. Instead, the narrow peaks in the blue trace are mostly due to mechanical thermal motion, i.e.
the thermal excitation of particular membrane normal modes, the most prominent of which are specified by the numbers atop of the associated peaks. Among all the modes, the following considers the mode (11), which presents itself as a split doublet due to the broken cylindrical symmetry of the membrane.

In Fourier space the equations for the displacement operators $\delta\tilde q_1$ and $\delta\tilde q_2$, associated with the two doublet modes, are
\vspace{2mm}
\begin{eqnarray}
       [\tcmea(\omega)]^{-1} \delta\tilde q_1 + \Sigma^\mathrm{eff}_{12}(\omega) \delta\tilde  q_2 
       			&=& \mathcal{N}_1^\mathrm{eff}(\omega)
				\\
       \Sigma^\mathrm{eff}_{21}(\omega) \delta\tilde q_1 + [\tcmeb(\omega)]^{-1} \delta\tilde q_2 
       			 &=& \mathcal{N}_2^\mathrm{eff}(\omega)\,,
\end{eqnarray}
\vspace{2mm}
where $\Sigma_{ij}^\mathrm{eff}(\omega)= \im g_{0i}\,g_{0j}\as^2
		\left\{\tcc^\mathrm{eff}(\omega) - [\tcc^\mathrm{eff}(-\omega)]^\ast\right\}$, with $\Sigma_{ij}^\mathrm{eff}(\omega) = \Sigma_{ji}^\mathrm{eff}(\omega)$ and $\Sigma_{ii}^\eff(\omega)$ commonly termed the (effective) ``self--energy''; we have further introduced $[\tcmei(\omega)]^{-1} = [\tcmi(\omega)]^{-1} + \Sigma_{jj}^\eff(\omega)$, with $[\tcmi(\omega)]^{-1} = [\wmj^2 - \omega^2 -\im \omega \gmj]/\wmj$ the ``bare'' mechanical suceptibility; finally, $ \mathcal{N}_j^\mathrm{eff}(\omega)= \tilde{\xi}_j(\omega) + g_{0j} \as \left[\tcce(\omega)\,\tilde n + [\tcce(-\omega)]^\ast\,\tilde n^\dag \right]$ is the noise reshaped by the effective cavity suceptibility.
As shown in~\cite{Genes2008} and~\cite{Shkarin2014}, the two mechanical modes with interaction mediated by the cavity field, $\Sigma_{12}(\omega) = \Sigma_{21}(\omega)$, can be recast in terms of ``bright'' and ``dark'' modes
\begin{eqnarray}
	\delta\tilde q_\mathrm{b} = \mu_1 \delta\tilde q_1 + \mu_{2}\delta\tilde q_2 
		\hspace{2cm}
	\delta\tilde q_\mathrm{d} = \mu_2 \delta\tilde q_1 - \mu_{1}\delta\tilde q_2 \,,
\end{eqnarray}
with $\mu_\mathrm{j} = g_\mathrm{0j}/\sqrt{g_{01}^2 + g_{02}^2}$. The system becomes 
\vspace{2mm}
\begin{eqnarray}
       [\tilde\chi_\mathrm{b}^\mathrm{eff}(\omega)]^{-1} &\delta\tilde q_\mathrm{b} 
		+ \Sigma^\mathrm{eff}_\mathrm{bd}(\omega) &\delta\tilde q_\mathrm{d}
       			= \mu_1\mathcal{N}_1^\mathrm{eff}(\omega) + \mu_2\mathcal{N}_2^\mathrm{eff}(\omega)
				\\
       \Sigma^\mathrm{eff}_\mathrm{db}(\omega) &\delta\tilde q_\mathrm{b} 
       		+[\tilde\chi_\mathrm{d}(\omega)]^{-1} &\delta\tilde q_\mathrm{d} 
       			 = \mu_2\mathcal{N}_1^\mathrm{eff}(\omega) - \mu_1\mathcal{N}_2^\mathrm{eff}(\omega)\,,
\end{eqnarray}
\vspace{2mm}
with $ [\tilde\chi_\mathrm{b}^\mathrm{eff}(\omega)]^{-1} =  [\tilde\chi_\mathrm{b}(\omega)]^{-1}+\Sigma^\mathrm{eff}_\mathrm{bb}(\omega)$, $ [\tilde\chi_\mathrm{b}(\omega)]^{-1} = \mu_1^2 [\tcma(\omega)]^{-1} + \mu_2^2 [\tcmb(\omega)]^{-1}$, $ [\tilde\chi_\mathrm{d}(\omega)]^{-1} = \mu_2^2 [\tcma(\omega)]^{-1} + \mu_1^2 [\tcmb(\omega)]^{-1}$, 
$\Sigma^\mathrm{eff}_\mathrm{bd}(\omega) = \Sigma^\mathrm{eff}_\mathrm{db}(\omega) = \mu_1\mu_2([\tcma(\omega)]^{-1}-[\tcmb(\omega)]^{-1})$, and $\Sigma^\mathrm{eff}_\mathrm{bb}(\omega) = \Sigma_{11}^\mathrm{eff}(\omega)+\Sigma_{22}^\mathrm{eff}(\omega)$. Note that $\Sigma^\mathrm{eff}_\mathrm{dd}(\omega)=0$, i.e. the ``dark'' mode susceptibility is not directly modified by the optomechanical interaction and it is coupled to the cavity mode only indirectly, through its coupling with the bright mode.

Finally, it is worthwile to define the resonance frequencies of the two hybrid modes as $\omega_\mathrm{b} = \mu_1^2 \, \omega_{\mathrm{m},1} + \mu_2^2 \, \omega_{\mathrm{m},2}$ and $\omega_\mathrm{d} = \mu_2^2 \, \omega_{\mathrm{m},1} + \mu_1^2 \, \omega_{\mathrm{m},2}$, which follow from the definitions of the corresponding mechanical suceptibilities. As a consequence of the feedback--enhanced optomechanical interaction, the ``bright'' mode then experiences an optical--spring effect, i.e. a shift in its resonance frequency, quantified by $\delta\omega_\mathrm{b} = \mathrm{Re}[\Sigma^\eff_\mathrm{bb}(\omega_\mathrm{b})]$. 

\begin{figure}[h!]
\centering
\includegraphics[width=0.75\textwidth]{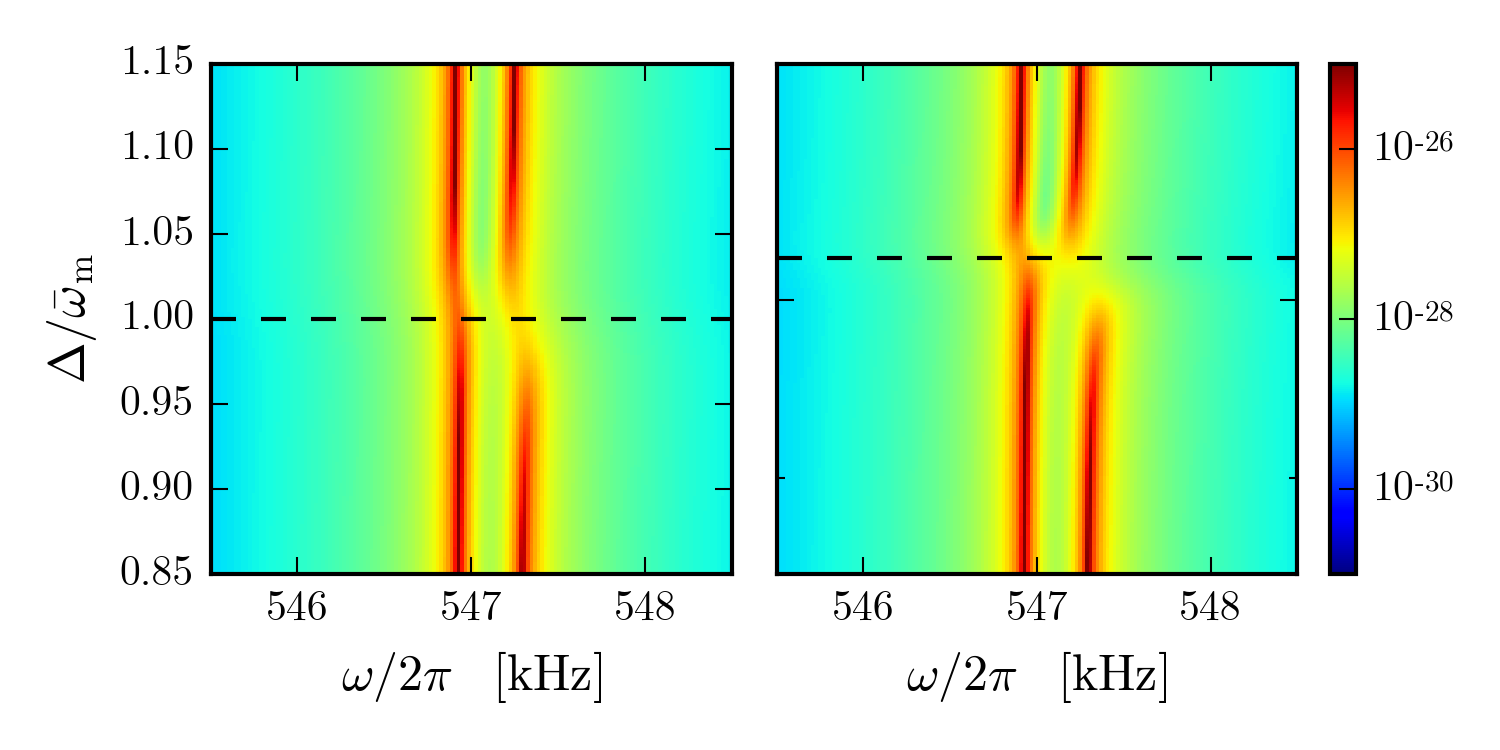}
\caption{Power noise displacement at the doublet (11) without (Left) and with (Right) feedback, as a function of the detuning normalised to the mean mechanical mode frequency $\bar\omega_\mathrm{m} = (\omega_\mathrm{m,1}+\omega_\mathrm{m,2})/2$. Cooling power set at $\mathcal{P} = \SI{74}{\micro\watt}$ and the feedback gain is $\mathcal{G} \sim 1$. The dashed black lines indicate the optimal detuning for cooling with feedback, $\Delta = 2\pi\times\SI{560}{\kilo\hertz}$, and that without, $\Delta = \bar\omega_{\mathrm{m}}$.}
\label{fig:Figure_9} 
\end{figure}

The theoretical displacement spectra of the doublet are shown in~\figurename~\ref{fig:Figure_9} as a function of the detuning normalised to the mean doublet frequency $\bar\omega_\mathrm{m} = (\omega_\mathrm{m,1}+\omega_\mathrm{m,2})/2$, for a cooling power $\mathcal{P} = \SI{74}{\micro\watt}$, without feedback and for positive feedback. 
The resonance frequency, mechanical damping rate and optomechanical coupling rate used are those 
obtained from the fit of the experimental data presented below: $\wml = 2\pi \times \SI{546.91}{\kilo\hertz}$, $\gml = 2\pi \times \SI{2.5}{\hertz}$ and $g_{0,1} = 2\pi \times \SI{.42}{\hertz}$ for the first mode and $\wmr = 2\pi \times \SI{547.26}{\kilo\hertz}$, $\gml = 2\pi \times \SI{3}{\hertz}$ and $g_{0,2} = 2\pi \times \SI{.67}{\hertz}$ for the second one. Dashed black lines indicate optimal detuning values for standard sideband cooling and sideband cooling with feedback--controlled light; the right panel shows an improvement of the effect for both modes. It is also visible that the lower--frequency mode is only weakly coupled to the optical field, becoming almost ``dark''. The higher--frequency mode, on the other hand, experiences significantly enhanced cooling and a pronounced optical--spring effect, testifying to the hybridisation of the two original normal modes. In fact, the hybridisation becomes considerable precisely because the feedback--modified resonance shift is large enough to fulfill $|\omega_\mathrm{b} + \delta\omega_\mathrm{b} - \omega_{\mathrm{d}}| \gg \max |\Sigma_\mathrm{db}(\omega)|$, i.e. for the separation between the bright and dark mode to be larger than the coupling between the two.

\subsection{Measured spectra and phonon occupancy}

\begin{figure}[h!]
\centering
\includegraphics[width=0.75\textwidth]{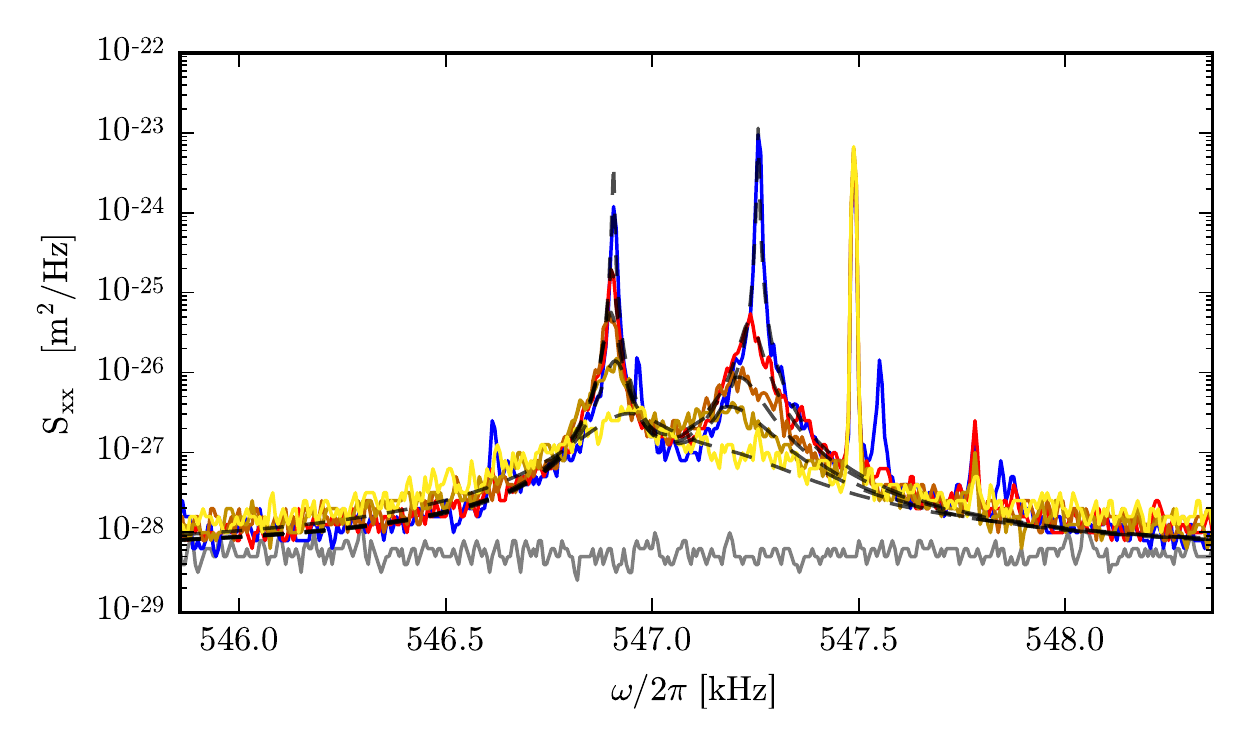}
\caption{Displacement spectral noise for different feedback gain and a fixed detuning of $\Delta = 2\pi\times\SI{560}{\kilo\hertz}$. The grey trace is the shot noise level; the blue trace represents thermal fluctuations of the mechanical mode doublet (11) in the absence of optomechanical effects, i.e. without sideband and feedback cooling. The narrow feature on the right is a tone used for calibrating $g_{0,j}$ and yielding $g_{0,1} = 2\pi \times \SI{.42}{\hertz}$ and $g_{0,2} = 2\pi \times \SI{.67}{\hertz}$. The estimated resonance frequencies and decay rates are $\omega_\mathrm{m,1} = 2\pi\times\SI{546.91}{\kilo\hertz}$ and $\gamma_\mathrm{m,1} = 2\pi\times\SI{2.5}{\hertz}$ for the left mode, and $\omega_\mathrm{m,2} = 2\pi\times\SI{547.26}{\kilo\hertz}$ and $\gamma_\mathrm{m,2} =2\pi\times\SI{3}{\hertz}$ for the right mode. The red trace shows the fluctuations of the mode reduced due to sideband cooling. Finally, traces from brown to yellow are taken with a fixed optical cooling rate, but turning on the feedback cooling and increasing the gain.}
\label{fig:Figure_10} 
\end{figure}

\begin{figure}[h!]
\centering
\includegraphics[width=0.75\textwidth]{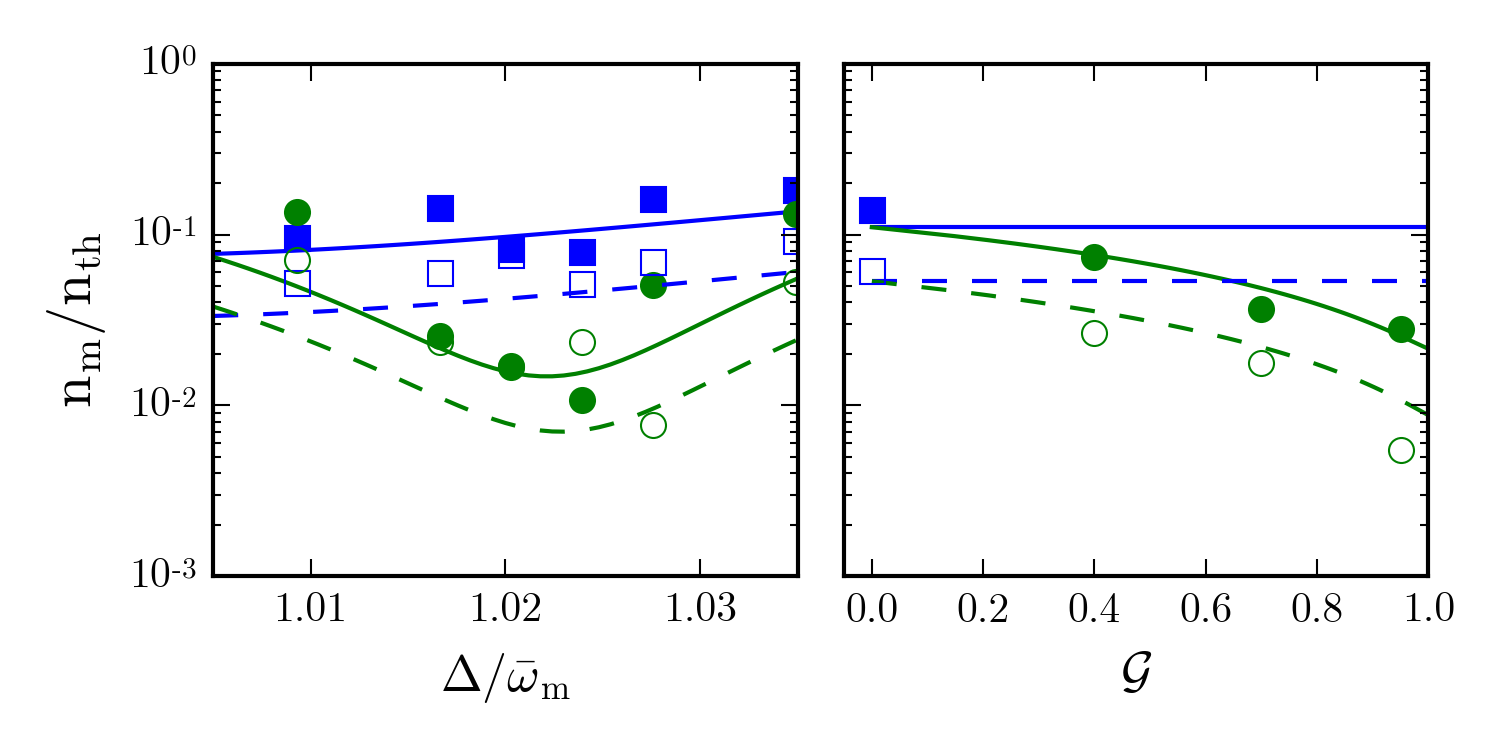}
\caption{Number of phonons, $\mathrm{n_m}$, of the doublet (11), normalised to the thermal phonon number $\mathrm{n_{th}}$ (essentially the same for the two modes) as a function of the detuning $\Delta$ normalised to $\bar\omega_\mathrm{m}$ (Left) and the feedback gain $\mathcal{G}$ (Right), without (blue lines) and with feedback (green lines). Full symbols and lines are the data and theoretical prediction for the right mode, whereas open symbols and dashed lines pertain to the left mode. The optimal cooling is obtained for a detuning $\Delta\sim 1.025\times\bar\omega_\mathrm{m} \sim 2\pi\times\SI{560}{\kilo\hertz}$, and for gain $\mathcal{G}\sim 1$. }
\label{fig:Figure_11} 
\end{figure}

A close--up of the mechanical doublet (11) displacement spectral density inferred from a homodyne measurement on the resonant probe beam is shown in~\figurename~\ref{fig:Figure_10}. 
Fitting the thermal fluctuations one infers the resonance frequencies and mechanical 
decay rates listed in the caption of \figurename~\ref{fig:Figure_10}, which were also used in the numerical simulations of \figurename~\ref{fig:Figure_9}. 
The narrow peak to the right of the doublet is an external calibration tone for estimating the two optomechanical couplings $g_{0,j}$ using the technique first shown in~\cite{Gorodetsky2010}. Then a detuned pump ($\Delta = 2\pi \times \SI{560}{\kilo\hertz}$, $\mathcal{P} = \SI{74}{\micro\watt}$) is turned on and the resulting optomechanical interaction sideband--cools both modes (red trace in \figurename~\ref{fig:Figure_10}). The cooling is improved by operating the feedback in the anti--squashing regime and increasing the gain, as testified by the brown to yellow traces. This improvement is properly quantified by examining the number of phonons $\mathrm{n_m}$ in the two modes, shown in~\figurename~\ref{fig:Figure_11} with and without feedback, normalised to the thermal occupancy $\mathrm{n_{th}}$. The corresponding reduction is obtained by numerical integration of the experimental spectra, assuming the equipartition theorem~\cite{Genes2008PRA}. The left panel inspects the dependence on the detuning, finding that the optimal value is $\Delta\sim 1.025\times\bar\omega_\mathrm{m}$, as indicated by the simulation in~\figurename~\ref{fig:Figure_9}. The right panel of~\figurename~\ref{fig:Figure_11} presents the phonon occupancy obtained by fixing the detuning at this value and increasing the gain, showing a clear reduction towards instability, i.e. $\mathcal{G}\sim 1$. In this regime feedback is able to further cool both vibrational modes, lowering the corresponding occupancies by one order of magnitude.

\section{Conclusion}

We have studied both theoretically and experimentally a multimode optomechanical system formed by a driven, feedback--controlled optical cavity mode, and two nearly degenerate mechanical modes of a thin SiN circular membrane~\cite{Serra2016}. The feedback loop is realised by measuring the intensity of the transmitted cavity output and using the detected signal to modulate the input light amplitude quadrature via the amplitude--modulation port of an AOM. The resulting in--loop optical field possesses phase--dependent fluctuations which can be either squashed (i.e., sub--shot noise) or anti--squashed (super--shot noise), depending upon the phase of the feedback gain, with the amount of either set by the gain amplitude. Usually in--loop fields are used in the squashing regime for stabilisation purposes, but here we have focused on the unusual anti--squashing regime, where the field fluctuations are increased and the feedback--controlled optical cavity system is close to instability. In this regime the cavity behaves as an effective cavity with a shifted resonance and a very narrow linewidth, which we have experimentally verified by measuring its response to a weak classical seed field, both for a quasi--resonant and for a very detuned beam. 

The in--loop field fluctuations can be tailored in order to control the dynamics of a generic system coupled to them, in the present case the two mechanical modes of a nearly degenerate doublet. We present the general theory of the modified dynamics of the two mechanical modes and in particular we show experimentally that in the anti--squashing regime the efficiency of simultaneous resolved--sideband cooling of the two modes can be significantly enhanced. By optimising the feedback parameters we decrease the resulting mechanical occupancy by one order of magnitude. 

The feedback technique illustrated here is very general and can be applied to improve the performance of every system coupled to the engineered in--loop field fluctuations. In particular it could be very useful in engineering and protecting against thermal decoherence effects on the quantum dynamics of the system. For example, as shown in Ref.~\cite{Rossi2017}, anti--squashed light allows a significant improvement in cooling mechanical resonators well below the quantum backaction limit, even past what has recently been achieved by injecting squeezed light~\cite{Clark2017}.

We acknowledge the support of the European Commission through the H2020-FETPROACT-2016 project n. 732894 ``HOT''.


\section{Bibliography}

\begin{thebibliography}{30}

\bibitem{WisemanMilburn2010}
Wiseman~H~M and Milburn~G~J
\newblock{2010 {\it Quantum Measurement and Control} (Cambridge: Cambridge University Press)}

\bibitem{Jacobs2014}
Jacobs~K
\newblock{2014 {\it Quantum Measurement Theory and its Applications} (Cambridge: Cambridge University Press)}

\bibitem{Taubman1995Intensity-feedb}
Taubman~M~S, Wiseman~H~M, McClelland~D~E and Bachor~H-A
\newblock{1995 {\it J. Opt. Soc. Am. B} {\bf 12} 1792--1800}

\bibitem{Wiseman1999Squashed-states}
Wiseman~H~M
\newblock{1999 {\it J. Opt. B: Quantum Semiclass. Opt.} {\bf 1} 459--463}

\bibitem{Shapiro1987Theory-of-light}
Shapiro~J~H, Saplakoglu~G, Ho~S~T, Kumar~P, Saleh~B~E~A and Teich~M~C 
\newblock{1987 {\it J. Opt. Soc. Am. B} {\bf 4} 1604--1620}

\bibitem{Wiseman1998In-Loop-Squeezi}
Wiseman~H~M
\newblock{1998 {\it Phys. Rev. Lett.} {\bf 81} 3840--3843}

\bibitem{Buchler1999Suppression-of-}
Buchler~B~C, Gray~M~B, Shaddock~D~A, Ralph~T~C and McClelland~D~E
\newblock{1999 {\it Opt. Lett.} {\bf 24} 259--261}

\bibitem{Sheard2005Experimental-de}
Sheard~B~S, Gray M~B, Slagmolen~J~J, Chow~J~H and McClelland~D~E
\newblock{2005 {\it IEEE J. Quan. Elect.} {\bf 41} 434--440}

\bibitem{Rossi2017}
Rossi~M, Kralj~N, Zippilli~S, Natali~R, Borrielli~A, Pandraud~G, Serra~E, Di~Giuseppe~G and Vitali~D
\newblock{arXiv:1704.04556}

\bibitem{AKM-RevMod}
Aspelmeyer~M, Kippenberg~T~J and Marquardt~F
\newblock{2014 {\it Rev. Mod. Phys.} {\bf86} 1391--1452} 

\bibitem{Wilson2015Measurement-bas}
Wilson~D~J, Sudhir~V,	Piro~N,	Schilling~R, Ghadimi~A and Kippenberg~T~J
\newblock{2015 {\it Nature} {\bf 524} 325--329}

\bibitem{McKenzie2002Experimental-De}
McKenzie~K, Shaddock~D~A, McClelland~D~E, Buchler~B~C and Lam~P~K
\newblock{2002 {\it Phys. Rev. Lett.} {\bf 88} 231102}

\bibitem{LIGO2013}
Aasi~J {\it et al.}
\newblock{2013 {\it Nat. Photon.} {\bf 7} 613--619}

\bibitem{Peano2015Intracavity-Squ}
Peano~V, Schwefel~H~G~L, Marquardt~C and Marquardt~F
\newblock{2015 {\it Phys. Rev. Lett.} {\bf 115} 243603}

\bibitem{Clark2016Observation-of-}
Clark~J~B, Lecocq~F, Simmonds~R~W, Aumentado~J and Teufel~J~D
\newblock{2016 {\it Nat. Phys.} {\bf 12} 683--687}

\bibitem{Schafermeier2016aa}
Sch{\"a}fermeier~C, Kerdoncuff~H, Hoff~U~B, Fu~H, Huck~A, Bilek~J, Harris~G~I, Bowen~W~P, Gehring~T and Andersen~U~L
\newblock{2016 {\it Nat. Commun.} {\bf 7} 13628}

\bibitem{Clark2017}
Clark~J~B, Lecocq~F, Simmonds~R~W, Aumentado~J and Teufel~J~D
\newblock{2017 {\it Nature} {\bf 541} 191--195}

\bibitem{Asjad2016Suppression-of-}
Asjad~M, Zippilli~S and Vitali~D
\newblock{2016 {\it Phys. Rev. A} {\bf94} 051801}

\bibitem{Teufel2011}
Teufel~J~D, Donner~T, Li~D, Harlow~J~W, Allman~M~S, Cicak~K, Sirois~A~J, Whittaker~J~D, Lehnert~K~W and Simmonds~R~W
\newblock{2011 {\it Nature} {\bf 475} 359--363}

\bibitem{Chan2011}
Chan~J, Mayer~Alegre~T~P, Safavi-Naeini~A~H, Hill~J~T, Krause~A, Gr\"{o}blacher~S, Aspelmeyer~M and Painter~O
\newblock{2011 {\it Nature} {\bf 478} 89--92}

\bibitem{Peterson2016Laser-Cooling-o}
Peterson~R~W, Purdy~T~P, Kampel~N~S, Andrews~R~W, Yu~P-L, Lehnert~K~W and Regal~C~A
\newblock{2016 {\it Phys. Rev. Lett.} {\bf116} 063601}

\bibitem{Genes2008}
Genes~C, Vitali~D and Tombesi~P
\newblock{2008 {\it New J. Phys.} {\bf10} 095009}

\bibitem{Shkarin2014}
Shkarin~A~B, Flowers-Jacobs~N~E, Hoch~S~W, Kashkanova~A~D, Deutsch~C, Reichel~J and Harris~J~G~E
\newblock{2014 {\it Phys. Rev. Lett.} {\bf112} 013602}

\bibitem{GardinerZoller}
Gardiner~C~W and Zoller~P
\newblock{2000 {\it Quantum Noise}, (Berlin: Springer)}

\bibitem{Drever1983}
Drever~R~W~P, Hall~J~L, Kowalski~F~V, Hough~J, Ford~G~M, Munley~A~J and Ward~H
\newblock{1983 {\it Appl. Phys. B} {\bf31} 97--105}

\bibitem{Yuen1983}
Yuen~H~P and Chan~V~W~S
\newblock{1983 {\it Optics Letters} {\bf8} 177--179}

\bibitem{Karuza2013}
Karuza~M, Biancofiore~C, Bawaj~M, Molinelli~C, Galassi~M, Natali~R, Tombesi~P, Di~Giuseppe~G and Vitali~D
\newblock{2013 {\it Phys. Rev. A} {\bf88} 013804}

\bibitem{Biancofiore2011}
Biancofiore~C, Karuza~M, Galassi~M, Natali~R, Tombesi~P, Di~Giuseppe~G and Vitali~D
\newblock{2011 {\it Phys. Rev. A} {\bf84} 033814}

\bibitem{Serra2016}
Serra~E, Bawaj~M, Borrielli~A, Di~Giuseppe~G, Forte~S, Kralj~N, Malossi~N, Marconi~L, Marin~F, Marino~F, Morana~B, Natali~R, Pandraud~G, Pontin~A, Prodi~G~A, Rossi~M, Sarro~P~M, Vitali~D and Bonaldi~M
\newblock{2016 {\it AIP Advances} {\bf6} 065004}

\bibitem{Gorodetsky2010}
Gorodetsky~M~L, Schliesser~A, Anetsberger~G, Deleglise~S and Kippenberg~T~J
\newblock{2010 {\it Opt. Express} {\bf18} 23236--23246}

\bibitem{Genes2008PRA}
Genes~C, Vitali~D, Tombesi~P, Gigan~S and Aspelmeyer~M
\newblock{2008 {\it Phys. Rev. A} {\bf77} 033804}

\end{thebibliography}
\end{document}